# The Two-Point Correlation Function of Rich Clusters of Galaxies: Results From An Extended APM Cluster Redshift Survey

G. B. Dalton[1], R. A. C. Croft[1], G. Efstathiou[1], W.J. Sutherland[1], S.J. Maddox[2] and M. Davis[3]

[1] *Department of Physics, University of Oxford, Keble Road, Oxford, OX1 3RH, UK.*
[2] *Royal Greenwich Observatory, Madingley Road, Cambridge, CB3 0EZ*
[3] *Departments of Physics and Astronomy, University of California, Berkeley, Ca 94720, USA.*

**ABSTRACT**

We present new estimates of the spatial two-point correlation function of rich clusters of galaxies selected from the APM Galaxy Survey. We have measured redshifts for a sample of 364 clusters out to a depth of $\sim 450\ h^{-1}$Mpc*. The clusters have a mean space density of $\bar{n} = 3.4 \times 10^{-5}\ h^3 \text{Mpc}^{-3}$. The two-point correlation function, $\xi_{cc}$, for this sample is equal to unity at a pair-separation of $r_0 = 14.3 \pm 1.75\ h^{-1}$Mpc ($2\sigma$ errors), consistent with our earlier results from a smaller sample. The new observations provide an accurate determination of the shape of $\xi_{cc}$ to pair-separations of about $50\ h^{-1}$Mpc. Our results show that $\xi_{cc}$ has a higher amplitude than expected according to the standard $\Omega = 1$ cold dark matter (CDM) model on spatial scales $2 \lesssim s \lesssim 50\ h^{-1}$Mpc, but are in good agreement with scale-invariant fluctuations in either a low density CDM model or a critical density universe made up of a mixture of hot and cold dark matter. Our results provide strong constraints on so called 'co-operative' models of galaxy formation in which the galaxy formation process introduces large-scale structure in the galaxy distribution.

## 1 INTRODUCTION

Rich clusters of galaxies have been used as tracers of large-scale structure by a number of authors over the last few years. Many of these studies have been based on Abell's (1958) cluster catalogue, which was compiled by visual inspection of uncalibrated photographic plates. However, there is considerable evidence that the clustering measured from Abell's catalogue is affected by inhomogeneities in the selection of clusters on the plane of the sky (Sutherland 1988, Soltan 1988, Efstathiou et al. 1992). As a result, several groups have attempted to produce more uniform samples by applying computer algorithms to select rich clusters of galaxies from calibrated photographic plates (Dalton et al. 1992, hereafter DEMS, Nichol et al. 1992) or by constructing X-ray flux limited samples (Romer et al. 1993, Nichol, Briel & Henry 1994). The results from these newer surveys give

$$\xi_{cc}(s) \approx (r_0/s)^2, \quad 13 \lesssim r_0 \lesssim 16\ h^{-1}\text{Mpc}, \quad (1)$$

*i.e.* the correlation length, $r_0$, is lower than the value $r_0 \lesssim 20\ h^{-1}$Mpc deduced from redshift surveys of Abell clusters (Bahcall & Soneira 1983, Klypin & Kopylov 1983, Postman, Huchra & Geller 1992), though Bahcall & West (1992) have argued that the discrepancy may be a consequence of a correlation between $r_0$ and cluster richness rather than inhomogeneities in the Abell catalogue. The computer-selected and X-ray selected redshift surveys have so far been limited to small samples of typically 100-200 clusters and so the shape of $\xi_{cc}$ is relatively poorly constrained, especially on scales $\gtrsim 15\ h^{-1}$Mpc where $\xi_{cc}$ is smaller than unity.

In this *Short Communication* we present the results of a new redshift survey of 364 rich clusters of galaxies selected from the APM Galaxy Survey (Maddox et al. 1990a, Maddox, Efstathiou & Sutherland 1990). This is a extension of our earlier redshift survey of 220 APM clusters described in DEMS and Dalton et al. (1994). The larger sample described here allows us to determine the shape of the $\xi_{cc}$ more accurately and to larger scales than in our earlier work. The space density of the cluster sample described here is comparable to that of our original survey (DEMS) and so in this paper we concentrate on a comparison of $\xi_{cc}$ with theoretical predictions rather than investigating the dependence of $\xi_{cc}$ with cluster richness. The richness dependence of $\xi_{cc}$ will be discussed in a future paper using a new redshifts for a deeper sample of particularly rich APM clusters.

---

* Where $h$ is Hubble's constant in units of 100km s$^{-1}$ Mpc$^{-1}$.

The new redshift survey is described in Section 2, and estimates of the two-point correlation function are presented in Section 3. In Section 4 we compare our results with the predictions of various CDM-like models, using N-body simulations as described by Croft & Efstathiou (1994a). We show that the new results allow us to draw strong conclusions on the parameters of acceptable CDM-like models. Our results also provide constraints on models where a physical process other than gravity modulates the efficiency of galaxy formation on large scales, and so changes the large-scale properties of the galaxy distribution. (Rees 1986, Babul & White 1991, Bower et al. 1993).

## 2  THE CLUSTER SAMPLES

DEMS have described an algorithm for selecting rich clusters of galaxies from the APM Survey. Here we describe some changes to the DEMS algorithm that were made to produce an approximately volume limited cluster sample to greater depths so allowing us to extend the DEMS redshift survey.

The depth of the cluster sample used by DEMS was limited by the magnitude range used to define the cluster richness. The cluster richness, $\mathcal{R}$, was defined by DEMS to be the weighted number of galaxies in the magnitude range $(m_X - 0.5, m_X + 1.0)$ above the mean background count in the range $(m_X - 0.5, m_X + 1.5)$, where $m_X$ is the magnitude of the galaxy for which the weighted count above background exceeded $X = \mathcal{R}/2$. The cluster catalogue defined in this way is limited by the point at which the faint end of the background slice equals the magnitude limit of the survey, which occurs when $m_X = 19.0$. We increased the depth of our new cluster catalogue by changing the background slice to $(m_X - 0.5, m_X + 1.0)$, and redefining $X$ to be $\mathcal{R}/2.1$[†] As a starting point for our new catalogue we used the same set of percolation centres as DEMS. A detailed description of the cluster finding algorithm and a discussion of the effects of changing the various selection parameters will be given in a future paper (see also Dalton 1992).

In the new cluster catalogue 117 clusters with $\mathcal{R} > 50$ and $m_X < 19.2$ had redshifts measured from our earlier redshift survey, Dalton et al. (1994). Dalton et al. (1994) list redshifts for an additional 20 clusters which have $\mathcal{R} > 50$ and $19.2 < m_X < 19.5$. This left 223 clusters without redshifts to form a complete sample of clusters with $\mathcal{R} > 50$ and $m_X < 19.2$. We observed typically three galaxies in the central regions of 154 clusters in four nights using the 4m telescope at Cerro Tololo Inter-American Observatory (CTIO).

---

[†] We shall use the symbols $m_X$ and $\mathcal{R}$ to refer to magnitude and richness estimates within the catalogue defined here; henceforth we will use the subscript DEMS to refer to the analogous quantities defined for the DEMS catalogue.

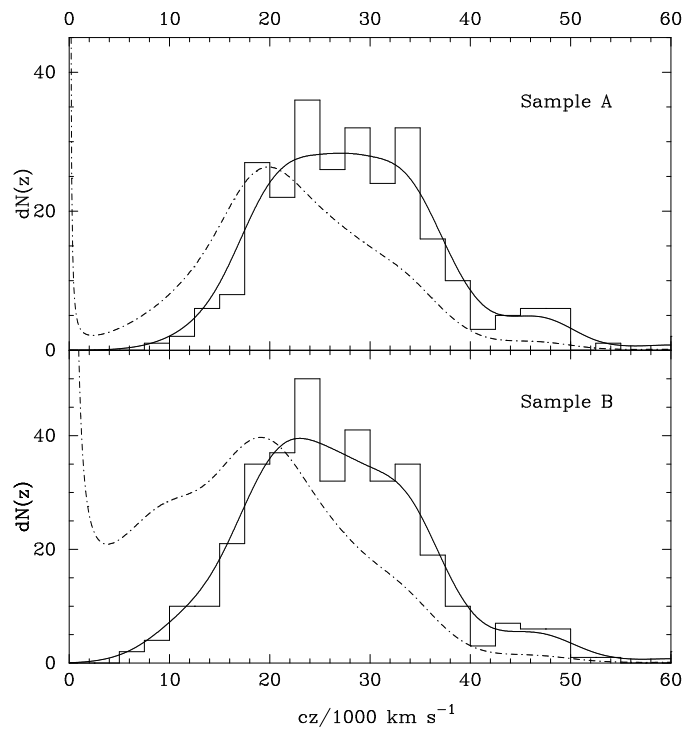

**Figure 1.** The redshift distributions of the two cluster samples discussed in the text. The solid line in each panel shows the smoothed distribution (see text). The dot-dashed lines show estimates of the selection functions (arbitrary normalisations) obtained by dividing the smoothed distribution by the volume element for an Einstein-de Sitter universe.

We applied the likelihood ratio test of Dalton et al. (1994) to each redshift measurement. We were unable to determine the redshift of 26 clusters because of poor quality spectra, or discrepancies in the galaxy redshifts none of which satisfied the likelihood ratio test as probable cluster members. We shall include the 20 clusters with previously measured redshifts, but with $m_X > 19.2$, and refer to this sample of 265 clusters as sample A. The mean space-density measured for this sample using the method described in Section 2.3 of Efstathiou et al. (1992) is $2.2 \times 10^{-5} \ h^3 \mathrm{Mpc}^{-3}$, similar to the space density of $2.4 \times 10^{-5} \ h^3 \mathrm{Mpc}^{-3}$ estimated for the DEMS $\mathcal{R}_{DEMS} > 20$ sample by Efstathiou et al. (1992).

We can increase the size of the redshift sample by transforming the richnesses of clusters in the DEMS sample with measured redshifts into the richness $\mathcal{R}$ defined here. Comparing the richnesses of clusters measured in both catalogues, we find the relation

$$\mathcal{R} = 24.8 + 1.2 \mathcal{R}_{DEMS}, \qquad (2)$$

with a scatter of 16 in $\mathcal{R}$. Transforming all of the DEMS richnesses to $\mathcal{R}$ using equation 2 and selecting clusters with $\mathcal{R} > 50$ leads to a sample of 364 clusters with measured redshifts, which we call sample B. The mean space density of clusters in sample B is approximately $3.4 \times 10^{-5} \ h^3 \mathrm{Mpc}^{-3}$. This sample is not limited strictly according to $\mathcal{R}$ as defined by the cluster selection. However it defines a statistical sample that is almost equivalent to a $\mathcal{R} \geq 50$ sample, because the scatter in the transformed richnesses is similar to the scatter intrinsic to the cluster finding algorithm (Dalton 1992).

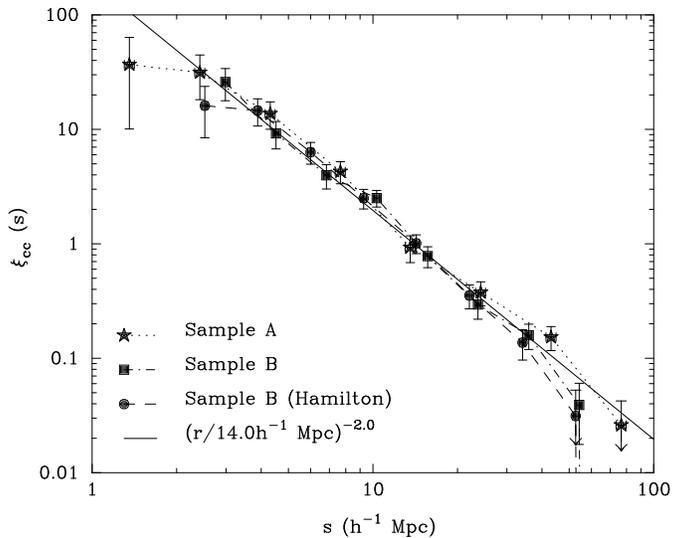

**Figure 2.** The correlation functions for samples A and B. The filled circles show the estimates obtained using Hamilton's (1993) estimator. The power law model $\xi_{cc} = (r/14\,h^{-1}\,\mathrm{Mpc})^{-2}$ which provided a good fit to the results of our earlier survey (Dalton et al. 1992) is shown for comparison.

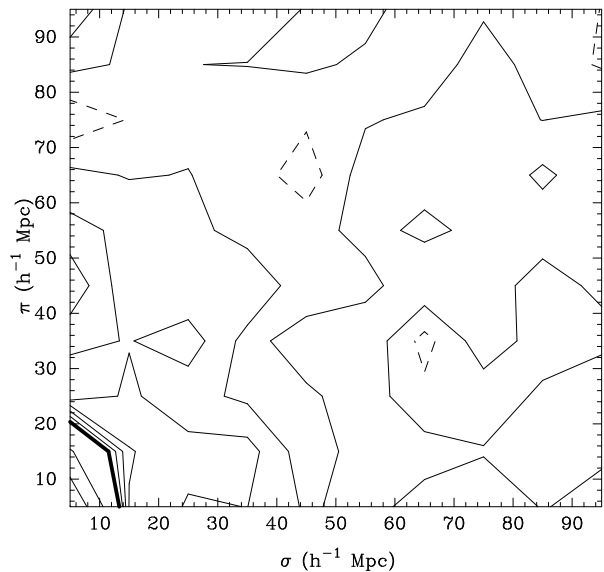

**Figure 3.** The correlation function $\xi_{cc}(\sigma, \pi)$ for sample B shown as a function of pair separations perpendicular to the line of sight ($\sigma$) and along the line of sight ($\pi$). Contour levels are at $\xi_{cc} = 3, 2, 1, 0.8, 0.6, 0.4, 0.2, 0., -0.2, -0.4$. The contour level at $\xi_{cc} = 1$ is shown by the heavy line; negative contours are plotted as dashed lines.

## 3 THE CLUSTER CORRELATION FUNCTION

We estimate the redshift-space correlation functions for the two samples using the estimator

$$\xi_{cc}(s) = 2f\frac{(DD)}{(DR)} - 1, \qquad (3)$$

where $f$ is the ratio of the number of random points to the number of clusters in the sample. In each case we use 20,000 points distributed within the survey boundaries and with the same redshift distributions as the smoothed distributions shown in Figure 1 which were obtained by convolving the observed redshift histograms with Gaussians of width $4000\,\mathrm{km\,s^{-1}}$. We estimated $\xi_{cc}$ using clusters with redshifts in the range in the range $5000$–$50000\,\mathrm{km\,s^{-1}}$. The correlation functions for the two samples are shown in Figure 2, together with estimates obtained from sample B using Hamilton's (1993) estimator:

$$\xi_{cc}(s) = 4\frac{(DD)\,(RR)}{(DR)^2} - 1, \qquad (4)$$

which is less affected by uncertainties in the selection function for $\xi_{cc} < 1$. In equations 3 and 4 the numerical factor accounts for the fact that we only count each $(DD)$ or $(RR)$ pair once. The error bars plotted in the Figure are computed from the formula $\delta\xi_{cc} = (1+\xi_{cc})/\sqrt{(DD)}$ which Croft & Efstathiou (1994a) have shown underestimates the variances in simulations of rich cluster catalogues in CDM-like models by a factor of between 1.3 and 1.7.

Fitting a power law

$$\xi_{cc}(s) = \left(\frac{s}{r_0}\right)^{-\gamma}$$

to the estimates of $\xi_{cc}$ over the range $1\,h^{-1}\,\mathrm{Mpc} \lesssim s \lesssim 40\,h^{-1}\,\mathrm{Mpc}$ for samples A and B plotted in Figure 2 gives:

$$\gamma = 2.14 \pm 0.19, \quad r_0 = 14.3^{+1.75}_{-1.5}, \quad \text{(Sample A)},$$
$$\gamma = 2.05 \pm 0.12, \quad r_0 = 14.3^{+1.75}_{-1.75}, \quad \text{(Sample B)},$$

where the error bars give an approximate indication of the 95% confidence intervals on the parameters of the fits. The agreement between samples A and B and the excellent agreement of the fits with those quoted by Efstathiou et al. (1992) shows that the correlation function is insensitive to changes in the cluster finding algorithm.

Figure 3 shows contour plots of the the correlation function for Sample B measured as a function of the pair separation along the line of sight, $\pi$, and perpendicular to the line of sight, $\sigma$. This Figure shows that $\xi_{cc}(\sigma, \pi)$ is nearly isotropic. There is no evidence of the large anisotropies measured in redshift surveys of Abell clusters by Efstathiou et al. (1992) in which correlations of order unity are measured to distances of $\pi \sim 100\,h^{-1}\,\mathrm{Mpc}$. The estimates of $\xi_{cc}(\sigma, \pi)$ for sample A are almost identical to those plotted in Figure 3 for sample B. The absence of anisotropies in the APM cluster catalogue provides strong evidence that the machine selected cluster samples are free of inhomogeneities on the plane of the sky.

## 4 COMPARISON WITH THEORY AND DISCUSSION

In Figure 4 we compare the estimates of $\xi_{cc}$ for sample B with theoretical predictions from dissipationless N-body simulations of CDM-like models (Croft & Efstathiou 1994a, Croft & Efstathiou 1994b). The three models shown are: (a) standard CDM, *i.e.* scale-invariant initial flucuations in an critical density universe with $h = 0.5$; (b) a low density CDM model (LCDM) with $\Omega = 0.2$, $h = 1$ and non-zero cosmological constant introduced to make the universe spatially flat; (c) a mixed dark matter (MDM) model with $h = 0.5$ in which CDM contributes $\Omega_{CDM} = 0.6$, light neutrinos contribute $\Omega_\nu = 0.3$ and baryons contribute $\Omega_B = 0.1$. The details of the simulations are given in Croft & Efstathiou (1994a) and Croft & Efstathiou (1994b).

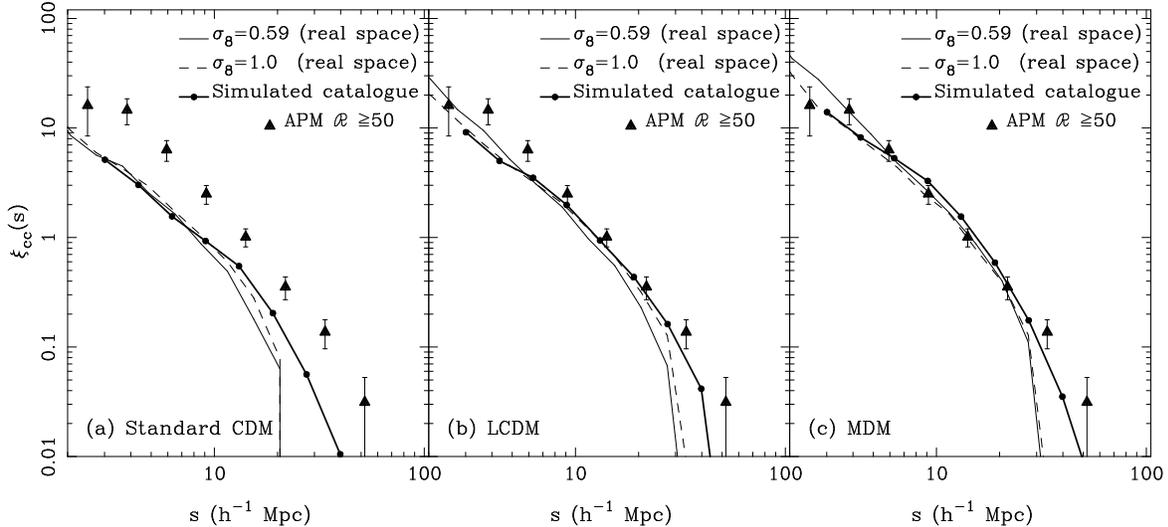

**Figure 4.** The data for sample B (filled circles in figure 2) compared with the average correlation functions of clusters selected from ten N-body simulations each of (a) Standard CDM, (b) a low density CDM model with non-zero cosmological constant, and (c) a mixed dark matter model (see text).

We locate rich clusters in real space in the simulations by first identifying cluster centres using a percolation algorithm and then computing the mass enclosed within a radius $r_c = 0.5\ h^{-1}$Mpc of the cluster centre-of-mass. The cluster catalogues are then ordered by mass and a lower mass limit is applied to the catalogue so that the mean space density of the clusters matches the observed space density ($3.4 \times 10^{-5}\ h^3$Mpc$^{-3}$ for sample B). Croft & Efstathiou (1994a) describe the algorithm used to select clusters in more detail and show that the two-point cluster correlation functions measured from the simulations are insensitive to the cluster identification radius $r_c$, to the amplitude of the mass fluctuations and, at the space-densities of interest here, to cluster richness.

Each panel in Figure 4 shows $\xi_{cc}$ determined from the simulations for two values of the amplitude of mass fluctuations as measured by $\sigma_8$, which gives the *rms* of the mass density fluctuations $\delta\rho/\rho$ measured in spheres of radius $8\ h^{-1}$Mpc. As shown by Croft & Efstathiou (1994a), the shape of $\xi_{cc}$ is remarkably insensitive to the value of $\sigma_8$. We note here that the COBE background radiation anisotropy measurements (Wright et al. 1994) require $\sigma_8 \approx 1$ for models (a) and (b) and $\sigma_8 \approx 0.67$ for model (c) (see Croft & Efstathiou 1994b) if the initial fluctuation spectrum is assumed to be scale invariant and temperature anisotropies from gravitational waves can be ignored. We have therefore used the simulation outputs for $\sigma_8 = 1$ for models (a) and (b) and $\sigma_8 = 0.6$ for model (c) to construct mock APM cluster catalogues in redshift space with the observed cluster redshift distribution and the APM Survey boundary. The two-point correlation functions measured from the simulated APM catalogues are plotted as the thick solid lines in Figure 4. The redshift-space cluster correlation functions for models (b) and (c) are almost indistinguishable from the real-space estimates, except on scales $\gtrsim 30\ h^{-1}$Mpc, where the redshift-space estimates are in better agreement with the observations. The difference between the redshift space and real space correlation function are more pronounced for model (a) with $\sigma_8 = 1$, though even here they are still relatively small.

Figure 4 shows that our new estimates of $\xi_{cc}$ lie higher than the predictions of the standard CDM model by $\approx 3\sigma$ *on all scales* plotted in the Figure (*i.e.* between $2\ h^{-1}$Mpc and $50\ h^{-1}$Mpc). The LCDM and MDM models give a much better match to the observed cluster correlations. In fact, the MDM model provides a slightly better fit to the observations than the $\Omega h = 0.2$ LCDM results plotted in Figure 4. However, an LCDM model with a lower value of $\Omega h$, may provide a more acceptable fit to the data (a numerical simulation of LCDM with $\Omega h = 0.15$ gave results in excellent agreement with the observations).

The LCDM and MDM models provide a good match to the galaxy correlations on scales $\gtrsim 5\ h^{-1}$Mpc measured from the APM Survey (Maddox et al. 1990b; Efstathiou, Bond & White 1992) if the galaxy two-point correlation function is related to the mass autocorrelation function by a linear biasing model, $\xi_{gg} = b^2 \xi_{\rho\rho}$, where $b$ is a constant. Some authors (e.g. Babul & White 1991) have argued that the galaxy formation process might introduce correlations on large scales, in which case the true relation between the galaxy and mass autocorrelation functions might be more complicated than assumed in the linear biasing model. For example, large-scale modulations in the galaxy distribution might arise if galaxies form with reduced or enhanced efficiency near to quasars (Rees 1986, Babul & White 1991, Efstathiou 1992, Bower et al. 1993).

The results shown in Figure 4 provide strong evidence against such models for the following reasons: (i) the correlation functions of rich clusters and galaxies have similar shapes but different amplitudes, as expected if clusters and galaxies are linearly biased tracers of the mass distribution; (ii) predictions of the rich cluster correlation function should be free of many of the uncertainties associated with the physics of galaxy formation if clusters form in regions of strong mass concentrations; (iii) the large differences between the observed form of $\xi_{cc}$ and the predictions of the standard CDM model, even on scales where $\xi_{cc}$ is greater than unity would seem difficult to explain by a spatial modulation of the efficiency of galaxy formation.

In summary, we have extended the APM cluster redshift survey to 364 rich clusters, sampling an effective volume of $\approx 10^7 \ h^{-3}\text{Mpc}^3$. The correlation function of rich clusters measured from this sample is in excellent agreement with our earlier results and is well approximated by a power law $\xi_{cc} = (r_0/s)^2$ with $r_0 = 14.3 \pm 1.75 \ h^{-1}\text{Mpc}$. N-body simulations of CDM-like models show that the standard CDM model fails to match the observations by a wide margin. The data are in good agreement with an MDM model with $\Omega_{CDM} \approx 0.6$ and $\Omega_\nu \approx 0.3$, or with a low-density variant of CDM with $\Omega h \sim 0.2$. Croft & Efstathiou (1994b) have argued that the peculiar velocities of clusters favour the MDM model over a low density CDM model, and analyses of galaxy streaming motions also favour $\Omega = 1$ universes (Kaiser et al. 1991, Dekel et al. 1993). A consistent picture of structure formation is therefore beginning to emerge in which $\Omega = 1$ and the mass distribution is more strongly clustered than expected according to the standard CDM model.

**Acknowledgements**

We thank the staff at CTIO for hospitality and efficient observing. This work was supported by grants from the UK Science and Engineering Research Council. RACC acknowledges receipt of a SERC studentship.